\documentclass[conference]{IEEEtran}
\IEEEoverridecommandlockouts
\usepackage{amsfonts}
\usepackage{times}
\usepackage{graphicx}
\usepackage{latexsym}
\usepackage{dsfont}
\usepackage{amssymb}
\usepackage{amsmath}
\usepackage{cite}
\usepackage{verbatim}

\newcommand{\figref}[1]{{Fig.}~\ref{#1}}

\def\bb0{{\mathbb{0}}}

\def\ba{{\mathbf{a}}}
\def\bb{{\mathbf{b}}}

\def\bh{{\mathbf{h}}}

\def\bm{{\mathbf{m}}}

\def\bp{{\mathbf{p}}}

\def\bz{{\mathbf{z}}}
\def\b0{{\mathbf{0}}}

\def\bA{{\mathbf{A}}}
\def\bB{{\mathbf{B}}}

\def\bD{{\mathbf{D}}}

\def\bZ{{\mathbf{Z}}}

\def\bbC{{\mathbb{C}}}

\def\bbE{{\mathbb{E}}}

\def\bbR{{\mathbb{R}}}

\def\cA{\mathcal{A}}

\def\cE{\mathcal{E}}
\def\cF{\mathcal{F}}

\def\cM{\mathcal{M}}
\def\cN{\mathcal{N}}
\def\cO{\mathcal{O}}

\def\cU{\mathcal{U}}
\def\cV{\mathcal{V}}

\def\sfg{{\mathsf{g}}}

\def\sf0{{\mathsf{0}}}

\usepackage{epstopdf}
\usepackage{enumerate}
\usepackage{algorithm}
\usepackage{amsmath}
\usepackage{float}
\usepackage{color}
\usepackage{makeidx}
\usepackage{bm}
\usepackage{cleveref}
\usepackage{url}
\usepackage{steinmetz}
\usepackage{varwidth}
\usepackage{soul}
\usepackage{booktabs}
\usepackage{subfigure}
\usepackage{balance}

\let\oldFootnote\footnote
\newcommand\nextToken\relax
\renewcommand\footnote[1]{
	\oldFootnote{#1}\futurelet\nextToken\isFootnote}
\newcommand\isFootnote{
	\ifx\footnote\nextToken\textsuperscript{,}\fi}

\DeclareMathOperator*{\argmax}{arg\,max}

\DeclareMathOperator\Arg{\mathrm{arg}}
\DeclareMathOperator\RE{\mathrm{Re}}
\DeclareMathOperator\IM{\mathrm{Im}}

\newcommand*{\J}{\jmath}
\newcommand*{\E}{\mathrm{e}}

\def \bEpsi{\boldsymbol{\Psi}}

\def \rm {\mathrm}
\def \bpsi{\boldsymbol{\psi}}

\begin{document}

	\title{Integrated Imaging and Communication with \\ Reconfigurable Intelligent Surfaces}

	\author{Hao Luo and Ahmed Alkhateeb \thanks{The authors are with Arizona State University (Email: h.luo, alkhateeb@asu.edu). This work is supported by the National Science Foundation under Grant No. 2229530.}
	}
	\maketitle

	\begin{abstract}	
	Reconfigurable intelligent surfaces, with their large number of antennas, offer an interesting opportunity for high spatial-resolution imaging. In this paper, we propose a novel RIS-aided integrated imaging and communication system that can reduce the RIS beam training overhead for communication by leveraging the imaging of the surrounding environment. In particular, using the RIS as a wireless imaging device, our system constructs the scene depth map of the environment, including the mobile user. Then, we develop a user detection algorithm that subtracts the background and extracts the mobile user attributes from the depth map. These attributes are then utilized to design the RIS interaction vector and the beam selection strategy with low overhead. Simulation results show that the proposed approach can achieve comparable beamforming gain to the optimal/exhaustive beam selection solution while requiring 1000 times less beam training overhead.
	\end{abstract}

	\section{Introduction} \label{sec:intro}
	Integrated sensing and communication (ISAC)~\cite{demirhan2023} has been identified as a key feature of future wireless systems. By incorporating these two functions into a single system, ISAC has the potential to enhance spectrum efficiency and reduce hardware cost and power consumption. Furthermore, ISAC has the potential to benefit both functionalities through communication-aided sensing and sensing-aided communication. Meanwhile, reconfigurable intelligent surfaces (RISs) have emerged as a promising approach to extend coverage and overcome blockages in both communication and sensing systems. The RIS can steer incident signals toward desired directions by adjusting the phase shifts of the passive reflecting elements. However, in RIS-assisted communication, the optimal configuration of the reflecting elements requires significant beam training overhead. Therefore, in this paper, we aim to develop an RIS-aided integrated imaging and communication system, where the RIS-based imaging of the surrounding environment can be used to mitigate the communication beam training overhead.

	Several studies have explored the use of RISs in ISAC systems~\cite{jiang2022a,he2022}. In~\cite{jiang2022a}, the RIS-aided dual-function radar and communication system is proposed, where the transmit precoding and passive reflection matrices are jointly optimized. In~\cite{he2022}, the authors study the joint design of the active beamforming of radar and the passive reflection matrices of two RISs in the communication radar coexistence system. So far, prior work has mainly focused on the interplay between communication and target detection, while other wireless sensing functionalities, e.g., imaging, have not been widely investigated.

	In this paper, we propose an RIS-aided integrated imaging and communication system, where the imaging-aided communication is achieved. Specifically, the contributions of this paper are organized as follows:
	\begin{itemize}
		\item We introduce a novel RIS-aided integrated imaging and communication system. The system leverages the high spatial dimensions of the RIS to perform wireless imaging and build a depth map of the surrounding environment. This depth map is then used to design the RIS interaction vector for communication with low overhead.
		\item We propose a user detection algorithm to extract the user position from an estimated depth map. Using the user position, we design the RIS interaction vector for communication. Then, we develop a beam selection strategy by considering a pre-defined RIS interaction codebook.
	\end{itemize}
	The simulation results demonstrate the capability of the designed solutions in achieving high beamforming gain and significantly reducing the RIS beam training overhead. This highlights the potential of the proposed RIS-aided integrated imaging and communication system.

	\textbf{Notation}: 
	$\bA$ is a matrix, $\ba$ is a vector, $a$ is a scalar. $\cA$ and $\boldsymbol{\cA}$ are sets of scalars and vectors. $\bA^T$, $\bA^H$, and $\bA^{\ast}$ are the transpose, Hermitian (conjugate transpose) and conjugate of $\bA$. $[\ba]_{n}$ is the $n^\rm{th}$ element of the vector $\ba$. $\rm{diag}(\ba)$ is a diagonal matrix with the entries of $\ba$ on its diagonal. $\bA \odot \bB$ is the Hadamard product of $\bA$ and $\bB$. $\cN(m,R)$ is a complex Gaussian random variable with mean $m$ and covariance $R$. $\RE(z)$, $\IM(z)$, and $\Arg(z)$ are the real part, the imaginary part, and the phase angle of the complex number $z$. $f(t) \ast g(t)$ is the continuous-time convolution of two signals $f(t)$ and $g(t)$. 

	\section{System Model} \label{sec:sys}
	\begin{figure*}[!t]
		\centering
		\subfigure[Imaging Stage]{\label{fig:sensing}\includegraphics[width=\columnwidth]{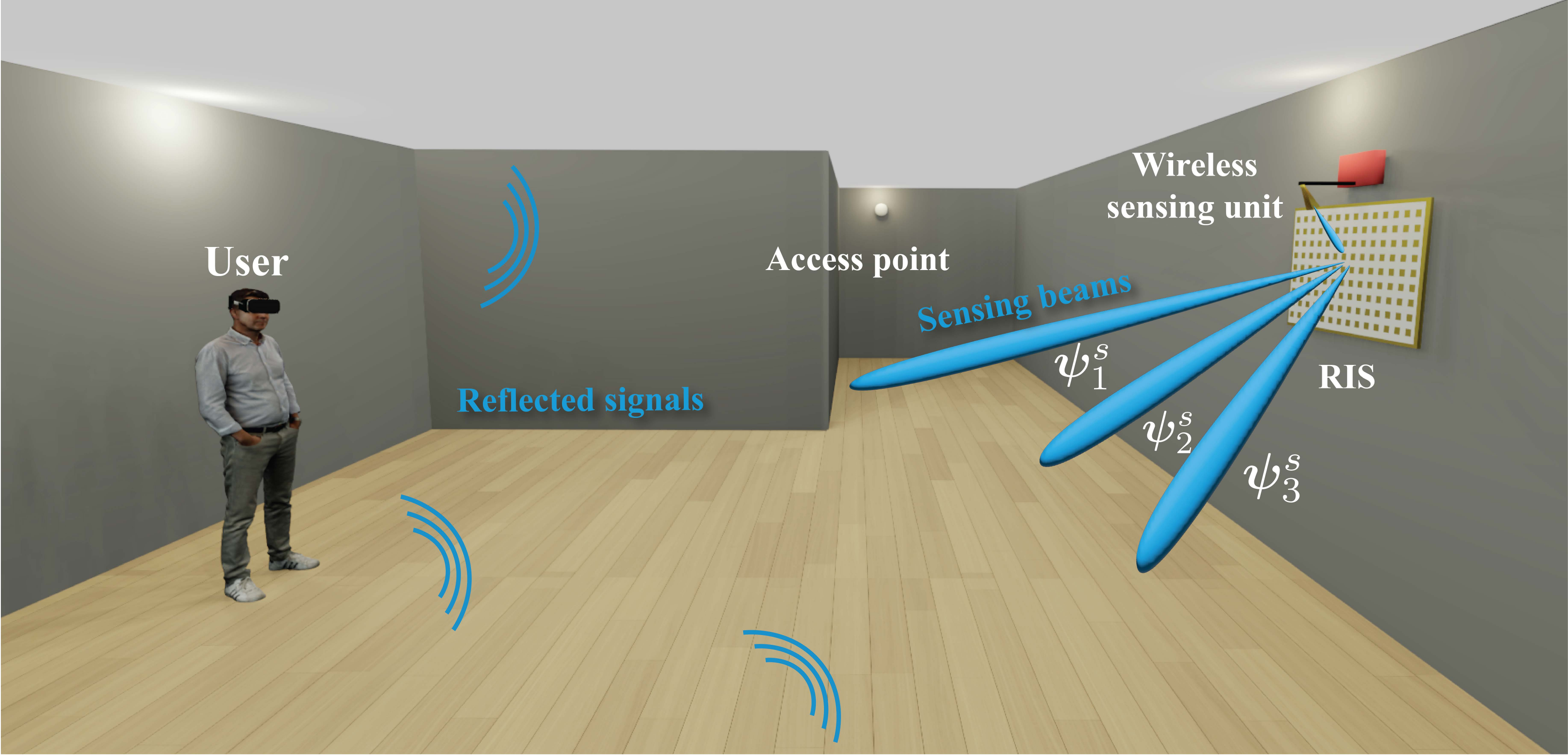}}
		\hfill
		\subfigure[Communication Stage]{\label{fig:comm}\includegraphics[width=\columnwidth]{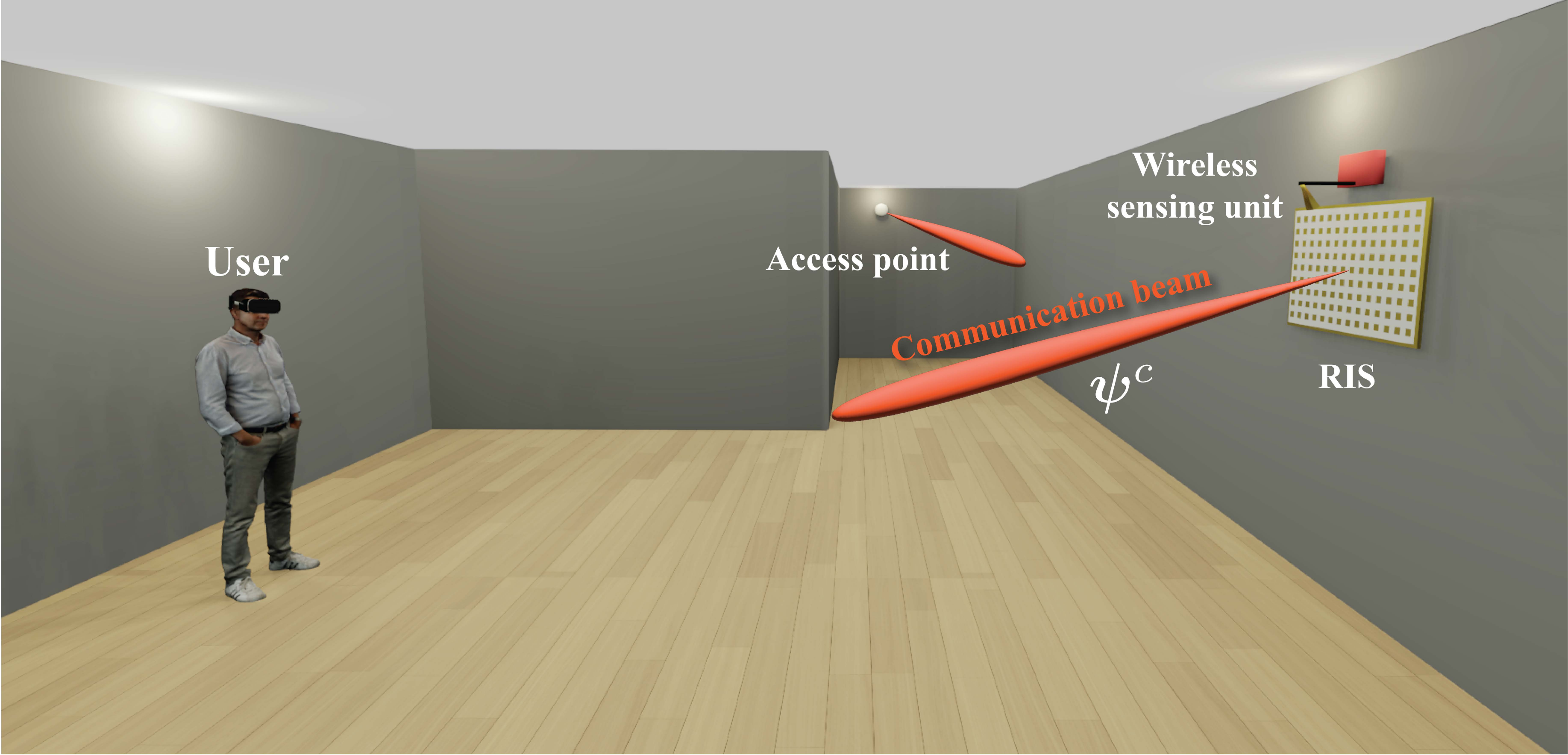}}
		\caption{This figure shows the integrated imaging and communication system. In the imaging/sensing stage, the wireless sensing unit transmits the sensing signals to the RIS via a feeding antenna. The sensing signals are reflected towards the environment by the RIS, which then reflects the backscattered/reflected signals back to the wireless sensing unit. The received signals are processed by the wireless sensing unit to construct a depth map of the environment, which enables the system to design the RIS interaction vector for communication.}
		\label{fig:system_model}
	\end{figure*}

	As illustrated in~\figref{fig:system_model}, we consider an RIS-aided integrated imaging and communication system, which consists of a mmWave access point (AP), a user, an RIS, and a mmWave wireless sensing unit. The mmWave access point, functioning as a transmitter, communicates with the user via the RIS. For simplicity, both the access point and the user equipment (UE) are assumed to have a single antenna structure. Also, it is assumed that there is no direct link between the access point and the user equipment. The mmWave wireless sensing unit is placed near the RIS and illuminates the non-line-of-sight (NLoS) area of the access point through the RIS. Following the design proposed in~\cite{taha2022}, the wireless sensing unit comprises a transmitter and a receiver that are connected to a shared single antenna through a self-isolation circuitry. The shared single antenna serves as a feeding antenna and transmits sensing signals to the RIS in order to achieve high spatial resolution.

	The RIS is assumed to be equipped with $N$ reconfigurable elements, and each element is modeled as a phase shifter. Denote the RIS interaction matrix for communication by $\bEpsi^{c} = \rm{diag}(\bpsi^{c}) \in \bbC^{N \times N}$, where $\bpsi^{c}=\left[\E^{\J\phi_{1}}, \dotsc, \E^{\J\phi_{N}}\right]^T$ is the interaction vector with unit modulus entries. The interaction matrix for sensing $\bEpsi^{s}$ can be defined similarly. In this paper, our objective is to detect the NLoS user from an estimated depth map and utilize it to devise the RIS interaction vector for communication. It is worth noting that the RIS-aided wireless sensing for depth estimation can be well-performed by an existing approach in the literature~\cite{taha2022}. Besides, we assume that the user is static during sensing and communication, and the scenario with the moving user is left to future work. Next, the communication and sensing models are described.

	\subsection{Communication Model}
	We assume the access point sends a complex symbol $x_c$ with the average power constraint $\bbE[|x_c|^2]={\cE_\rm{c}}$, and the RIS reflects the incident signal with the interaction vector $\bpsi^{c}$. Then, the received signal can be expressed as
	\begin{equation}
		y_c = \bh_{R}^{T} \bEpsi^{c} \bh_{T} x_c + w_c = (\bh_{R} \odot \bh_{T})^{T} \bpsi^{c} x_c + w_c,
	\end{equation}
	where $\bh_T \in \bbC^{N}$ is the channel between the AP and the RIS, and $\bh_R \in \bbC^{N}$ is the channel between the RIS and the user. $w_c \sim \cN(0,\sigma_{w}^2)  \in \bbC$ is the received noise. The channel $\bh_T$ can be defined as
	\begin{equation}
		\bh_T=\sum_{\ell=1}^{L} \alpha_{\ell} \ba(\theta_{\ell}^\rm{az}, \theta_{\ell}^\rm{ze}),
	\end{equation}
	where $\alpha_{\ell}$ is the complex-valued channel gain of path $\ell$ and $\ba(.)$ is the far-field transmit/receive array response vector of the RIS. $\theta_{\ell}^{\rm{az}}, \theta_{\ell}^{\rm{ze}}$ denote the azimuth and zenith angles of arrival, relative to the RIS. The channel $\bh_R$ can be defined similarly as $\bh_T$.

	\subsection{Sensing Model}
	For the sensing model, we adopt a wideband FMCW radar transceiver with a complex-baseband architecture~\cite{ramasubramanian2017,li2021}. The FMCW transmit signal is referred to as a radar frame, which contains a sequence of $M_{\rm{chirp}}$ chirps with a repetition interval of $T_{\rm{PRI}}$. A single linear chirp signal $a_\rm{BP}(t)$ can be written as 
	\begin{equation}
		a_\rm{BP}(t) =
		\begin{cases}
			\cos\left(2\pi f_{0}t + \pi S t^2\right) & 0\leq t \leq T_{\rm{active}},\\
			0 & \text{otherwise},
		\end{cases}
	\end{equation}
	where $f_{0}$ is the starting chirp frequency, and $T_{\rm{active}}$ is the duration of the chirp signal. $S=\rm{BW}/T_{\rm{active}}$ is the slope of the linear chirp with the bandwidth $\rm{BW}$. Then, the transmit signal of a radar frame $x_{s}^{\rm{BP}}(t)$ can be formulated as
	\begin{align}
		x_{s}^{\rm{BP}}(t) &=\sqrt{\cE_\rm{s}} \, \sum_{v=0}^{M_\rm{chirp}-1} a_\rm{BP}(t-vT_{\rm{PRI}})  \\
		& = \RE\left( x_s(t) \, \E^{\J 2\pi f_0 t } \right), t\in\bbR_{\geq 0},
	\end{align}
	where $\cE_\rm{s}$ is the transmit power gain and $x_s(t)\in \bbC$ is the complex-valued lowpass-equivalent transmit signal.

	The received bandpass signal $y_{s}^{\rm{BP}}(t) = \RE(y_s(t) \E^{\J 2\pi f_0 t })$ can be defined by the lowpass-equivalent received signal $y_s(t) \in \bbC$, which can be written as
	\begin{align}
		y_s(t) &= g(t) \ast x_s(t) + w_s(t) \\ 
		&= \sum_{k=1}^{K} \sum_{\ell=1}^{L_k} \sfg_{k,\ell} x_s(t - \xi_{k,\ell}(t) ) + w_s(t),
	\end{align}
	where $g(t)$ is the complex-valued lowpass-equivalent channel, and $w_s(t) \sim \cN(0,\sigma_{w}^2)  \in \bbC$ is the received noise at the wireless sensing unit.
	$K$ is the number of targets in the environment, and $L_k$ is the number of channel paths interacting with $k^{\rm{th}}$ target.
	$\sfg_{k, \ell} \in \bbC$ is the complex-valued channel gain of the $\ell^{\rm{th}}$ path of the $k^{\rm{th}}$ target.
	$\xi_{k, \ell} = R_{k,\ell}/\varsigma$ is the propagation delay, where $\varsigma$ is the speed of light, and $R_{k, \ell}$ is the propagation distance traveled by the $\ell^{\rm{th}}$ path of the $k^{\rm{th}}$ target.
	For further details of the channel model, please refer to~\cite{taha2022}.

	At the radar receiver, the received bandpass signal $y_{s}^{\rm{BP}}(t)$ is first passed through a quadrature mixer, and mixed with two versions of the transmit bandpass signal $x_{s}^{\rm{BP}}(t)$, one with a $-90^\circ$ phase shift.
	Then, low-pass filters and analog-to-digital converters (ADCs) are applied to the outputs of the mixer to generate the in-phase signal $I[u,v]$ and the quadrature-phase signal $Q[u,v]$, for the ADC sample $u \in \cU, \cU=\left\{0,1,\ldots,\left(M_\rm{sample}-1\right)\right\}$, and for the chirp $v \in \cV, \cV=\left\{0,1,\ldots,\left(M_\rm{chirp}-1\right)\right\}$. $M_\rm{sample}$ is the number of ADC samples per chirp.
	Let $F_S$ denote the ADC sampling frequency.
	The in-phase and quadrature-phase signals are sampled at time $t=uT_S+vT_\rm{PRI}$, $T_S=1/F_S$.
	Finally, the received baseband digital signal $z[u,v] = I[u,v] + \J \, Q[u,v]$ can be formulated as
	\begin{equation}
		z[u,v] = \sum_{k=1}^{K} \sum_{\ell=1}^{L_k} \sqrt{\rho_{k,\ell}} \, \E^{- \J \vartheta_{k,\ell}} \, \E^{+ \J \, \Xi_{k,\ell}} +  w_s[u,v] \E^{\J \chi[u]},
	\end{equation}
	Where $\chi[u] = 2\pi f_{0}t_{\rm{fast}} + \pi S t_{\rm{fast}}^2 $ and $t_{\rm{fast}}=uT_{\rm{S}}$. The channel path received power and phase are $\rho_{k,\ell}= \cE_\rm{s} |\sfg_{k,\ell}|^{2}$ and $\vartheta_{k,\ell}=\Arg\left( \sfg_{k,\ell}\right)$, respectively. The phase term, $\!\! \Xi_{k,\ell} \!  = \!  2\pi  \! \left( f_0 \xi_{k,\ell}  +  S t_{\rm{fast}} \xi_{k,\ell} -  \tfrac{S}{2} \xi_{k,\ell}^{2} \right) \!$, contains the range information of the target.

	\section{Problem Definition} \label{sec:prob}
	In this paper, we aim to position the user in the NLoS area and design the RIS interaction vector for communication based on the depth map estimated by the RIS-aided wireless sensing unit. For the sensing purpose, we adopt the design of the RIS sensing codebook in~\cite{taha2022}, denoted by $\boldsymbol{\cF}_s=\{\bpsi_m^{s}: m \in \cM_s, \cM_s = \{0,\ldots,M_s-1\}\}$. The codebook builds a rectangular sensing grid of reflected directions $\cO = \{(\theta_m^\rm{az}, \theta_m^\rm{ze})_{m=0}^{M_s-1}\}$ with $M_h$ and $M_v$ beams in the horizontal and vertical dimensions, i.e., $M_s = M_v M_h$. Then, the wireless sensing unit can construct a depth map of resolution $M_h$ pixels wide and $M_v$ pixels high. During the sensing process, the wireless sensing unit sweeps over the codebook, where each RIS beam participates in the transmission and reception of a single chirp. For the $m^{\rm{th}}$ RIS interaction vector, the received baseband digital signal can be expressed as
	\begin{multline} \label{eq:z}
		z[u,m] = 
		\\ \underbrace{\sum_{k=1}^{K} \sum_{\ell=1}^{L_k} \sqrt{\rho_{k,\ell}[m]} \E^{- \J \vartheta_{k,\ell}[m]} \E^{+ \J \, \Xi_{k,\ell}}}_\text{Received signal} + \underbrace{w_s[u,m] \E^{\J \chi[u]}}_\text{Noise} \, .
	\end{multline}
	By stacking the received $M_\rm{sample}$ ADC samples of each chirp, the received sensing signal matrix $\bZ \in \bbC^{M_{\rm{sample}}\times M_s}$ can be constructed as
	\begin{equation}
		\bZ=\left[\bz[0], \bz[1], \dotsc, \bz[M_s-1] \right],
	\end{equation}
	\begin{equation}
		\bz[m] = \left[ z[0,m],  \dotsc, z[M_\rm{sample}-1,m] \right]^T.
	\end{equation}
	To obtain the depth map, we adopt the approach in~\cite{taha2022} to process the received signals of the beams and estimate the depth value of each pixel. The estimated depth map $\bD_\rm{map} \in \bbR^{M_v \times M_h}$ can be formulated as 
	\begin{equation}
		\bD_{\rm{map}}=\bp_d(\bZ; \boldsymbol{\cF}_s),
	\end{equation}
	where $\bp_d(.)$ denote the depth map estimation function.

	Next, our objective is to develop a user detection function $\bp_u(.)$ that is able to estimate the azimuth and zenith angles towards the user, relative to the RIS, given by
	\begin{equation}
		(\tilde{\theta}_{\rm{UE}}^\rm{az}, \tilde{\theta}_{\rm{UE}}^\rm{ze}) =\bp_u(\bD_{\rm{map}}).
	\end{equation}
	With the acquired angle information of the user, we then propose to design an RIS interaction vector to maximize the received SNR at the UE. Given the communication model described in Section~\ref{sec:sys}, the received SNR can be written as
	\begin{equation} \label{eq:snr}
		\rm{SNR} = \frac{\cE_\rm{c}}{\sigma_w^2}|(\bh_{R} \odot \bh_{T})^{T} \bpsi^{c}|^2.
	\end{equation}
	Thus, the optimal RIS interaction vector for communication can be obtained by solving the following optimization problem
	\begin{gather} \label{eq:prob1}
		\begin{aligned}
			\max_{\bpsi^c} \quad & |(\bh_{R} \odot \bh_{T})^{T} \bpsi^{c}| \\
			\textrm{s.t.} \quad & |[ \bpsi^c ]_n|=1, \forall n\in \left\{1,\ldots,N\right\}.
		\end{aligned}
	\end{gather}
	If we assume that the RIS has a pre-determined communication codebook $\boldsymbol{\cF}_c, |\boldsymbol{\cF}_c|=M_c$, e.g., a beamsteering/DFT codebook, the objective of the optimization problem becomes finding the optimal beam index in the codebook, which can be formulated as follows:
	\begin{gather}
		\begin{aligned}
			m^\star = \argmax_{\psi_m^c \in \bm{\cF}_c} \quad & |(\bh_{R} \odot \bh_{T})^{T} \bpsi_m^{c}|
		\end{aligned}
	\end{gather}
	where the solution can be obtained by an exhaustive search, i.e., beam sweeping, over all the beams in the communication codebook. However, the large codebook would result in a significant beam training overhead. Therefore, we propose to leverage the sensing to reduce the number of trials, which will be presented in the next section. 

	\section{Proposed Solutions} \label{sec:sol}
	\begin{figure*}[!t]
		\centering
		\includegraphics[width=1.99\columnwidth]{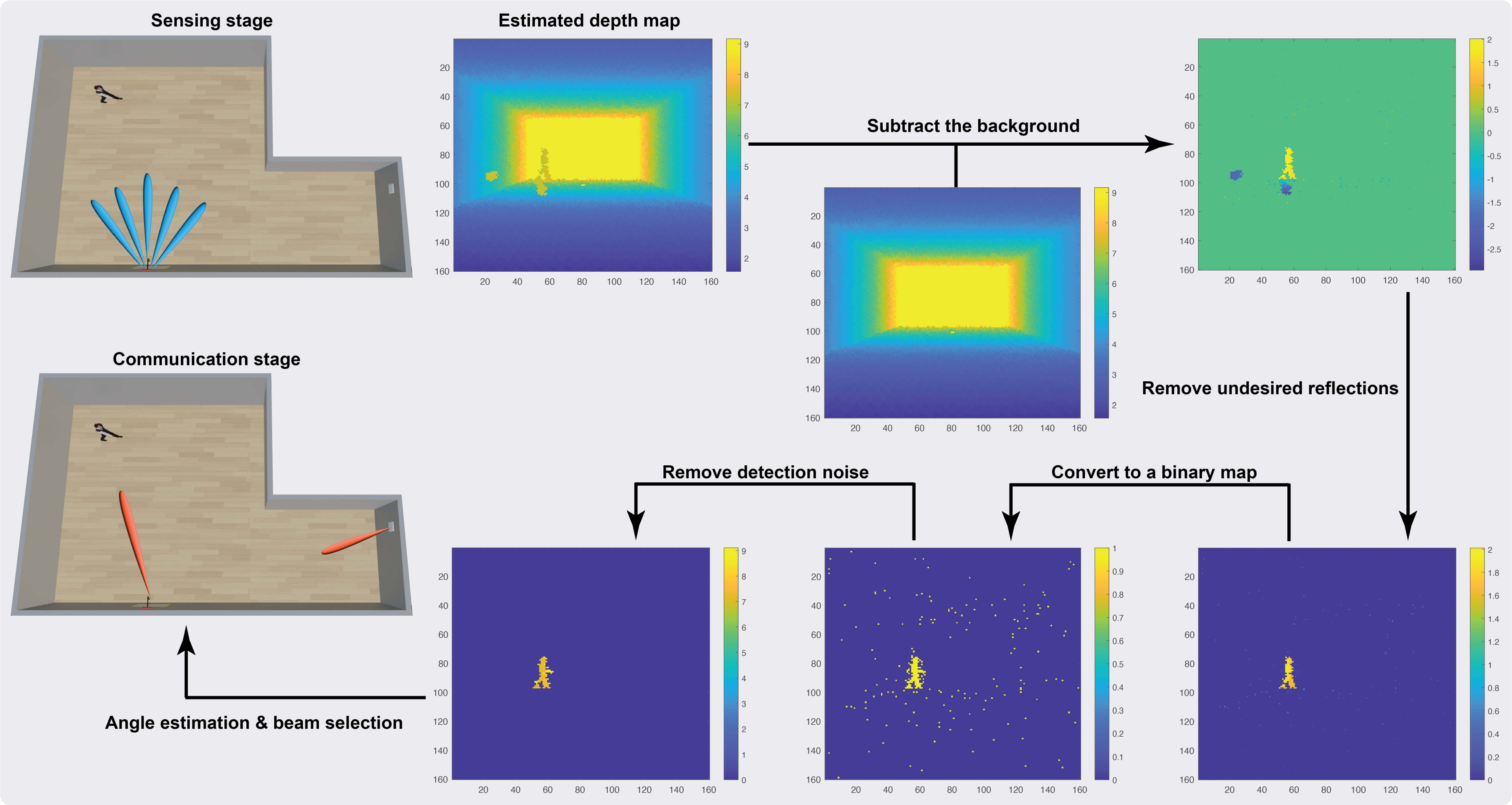}
		\caption{This figure presents the operation flow of the proposed image-aided communication solution. To begin, the estimated depth map undergoes background subtraction, followed by the elimination of undesired reflections and sensing noise. The pixel coordinates and the corresponding angles of the user can then be obtained from the processed depth map. Finally, the beam selection strategy is performed based on the estimated angles of the user.}
		\label{fig:solutions}
	\end{figure*}

	In this section, we first introduce the proposed approach to positioning the user in the depth map. Next, we describe the RIS interaction vector design for communication. The overall procedure of the proposed solutions are shown in \figref{fig:solutions}.

	\subsection{User Detection with Background Subtraction} \label{sec:user_detect}
	The main concept of user detection is to extract useful user information from the depth map, e.g., the user's pixels, which can further be leveraged for communication purposes. To achieve this goal, we design the following key steps.

	\textbf{Background Subtraction:}
	For user detection, the objective is to identify which pixels in the depth map belong to the user. This can be done by subtracting the background, which effectively removes most of the unwanted regions in the depth map. Such background depth map can be, for instance, estimated during the offline stage where there is no user in the scene. Let $\bD_\rm{map}^{b}$ denote the background depth map. The background-subtracted depth map $\bD_\rm{map}^{bs}$ can be obtained by
	\begin{equation}
		\bD_\rm{map}^{bs} = \bD_\rm{map}^{b} - \bD_\rm{map}.
	\end{equation}
	Note that, we subtract the estimated depth map $\bD_\rm{map}$ from the background $\bD_\rm{map}^{b}$ to make the user's pixels have positive values in the background-subtracted depth map $\bD_\rm{map}^{bs}$.

	\textbf{Removal of Undesired Reflections:}
	Ideally, each beam in the RIS sensing codebook should detect the range/depth of the single-backscattering path of its pointing direction. However, some undesired reflections provide larger channel path gain than the single-backscattering paths, resulting in higher estimated depth values. As a result, there are some regions with negative values in the background-subtracted depth map $\bD_\rm{map}^{bs}$ as shown in \figref{fig:solutions}. To remove these undesired reflections, we can set the negative values in $\bD_\rm{map}^{bs}$ to zero. This has no impact on the detected user's pixels from the previous step.

	\textbf{Elimination of Detection Noise:}
	Due to the time-varying noise in the sensing process, the background-subtracted depth map suffers from the detection noise, which can be better observed after converting to a binary map. To resolve this, we first identify the pixel coordinates with positive values in the background-subtracted depth map. Next, since the noise pixels are sparse, we leverage a density-based clustering algorithm, DBSCAN~\cite{ester1996}, to separate the user from the pixels. In the single-user scenario, we choose the cluster with the highest number of elements as the detected user. To determine the azimuth and zenith angles towards the user, we calculate the rounded mean of the user's pixel coordinates, denoted by $(x_u, y_u)$, representing the detected user. Given that each pixel is estimated by a beam of a pre-defined reflected direction, we can obtain the angles towards the user using the corresponding pixel coordinate, i.e., $\tilde{\theta}_{\rm{UE}}^\rm{az}=\theta_{m_u}^\rm{az}$, $\tilde{\theta}_{\rm{UE}}^\rm{ze}=\theta_{m_u}^\rm{ze}$, $m_u=x_u+M_hy_u$, where $(\theta_{m_u}^\rm{az}, \theta_{m_u}^\rm{ze})\in \cO$.

	\subsection{RIS Interaction Vector Design for Communication}
	To design the RIS interaction vector for communication, we can first decompose it into the AP-side and the UE-side beams~\cite{jiang2022b} as shown by
	\begin{equation}
		\bpsi^c = \bpsi_\rm{AP}^c \odot \bpsi_\rm{UE}^c,
	\end{equation}
	where $\bpsi_\rm{AP}^c$ and $\bpsi_\rm{UE}^c$ denote the AP-side and UE-side RIS interaction vectors. Accordingly, the optimization problem in~\eqref{eq:prob1} can be rewritten as
	\begin{gather} \label{eq:re_prob1}
		\begin{aligned}
			\max_{\bpsi_\rm{AP}^c, \bpsi_\rm{UE}^c} \quad & |(\bh_{T} \odot \bpsi_\rm{AP}^c)^{T} (\bh_{R} \odot \bpsi_\rm{UE}^{c})| \\
			\textrm{s.t.} \quad & |[ \bpsi_\rm{AP}^c ]_n|=1, \\
			\quad & |[ \bpsi_\rm{UE}^c ]_n|=1, \forall n\in \left\{1,\ldots,N\right\}.
		\end{aligned}
	\end{gather}
	In mmWave communications, the line-of-sight (LoS) path provides the dominant channel gain.
	Assuming the locations of the RIS and the AP are known in advance, we can design the AP-side beam to match the LoS propagation path. Thus, the AP-side RIS interaction vector can be expressed as 
	\begin{equation}
		\bpsi_\rm{AP}^{c} = \ba^{*}(\theta_\rm{AP}^\rm{az}, \theta_\rm{AP}^\rm{ze}),
	\end{equation}
	where $\ba(.)$ is the far-field RIS array response vector, and $\theta_\rm{AP}^\rm{az}$, $\theta_\rm{AP}^\rm{ze}$ denote the azimuth and zenith angles towards the AP, relative to the RIS. Similarly, the UE-side RIS interaction vector $\bpsi_\rm{UE}^{c}$ can be designed to focus on the LoS path between the RIS and the UE. With the estimated angles towards the user $\tilde{\theta}_\rm{UE}^\rm{az}, \tilde{\theta}_\rm{UE}^\rm{ze}$, the RIS interaction vector can be written as 
	\begin{equation}
		\tilde{\bpsi}^{c} = (\ba(\theta_\rm{AP}^\rm{az}, \theta_\rm{AP}^\rm{ze}) \odot \ba(\tilde{\theta}_\rm{UE}^\rm{az}, \tilde{\theta}_\rm{UE}^\rm{ze}))^{*}. 
	\end{equation}
	However, this approach requires accurate angle estimation, which may not be feasible in all circumstances due to the sensing noise. Therefore, we then propose an RIS beam selection scheme based on a pre-defined codebook, where a set of candidate beams can be found.

	Considering the codebook constraint, we can determine the beam index by calculating the similarities between the RIS interaction vector $\tilde{\bpsi}^{c}$ and the beams in the codebook. Then, the selected beam index $\bar{m}$ is represented as
	\begin{equation} \label{eq:sim}
		\tilde{m}=\argmax_{\psi_m^c \in \bm{\cF}_c}\ |(\tilde{\bpsi}^{c})^H  \bpsi_m^{c}|
	\end{equation}
	Note that, we can find a set of candidate beams by sorting the codebook based on the calculated similarities in \eqref{eq:sim}.

	\section{Simulation Results} \label{sec:sim}
	In this section, we evaluate the performance of the proposed solutions for the integrated imaging and communication system. We first describe the adopted simulation framework and then present the performance of the RIS beam selection in an indoor scenario.
	\begin{figure}[!t]
		\centering
		\includegraphics[width=1\columnwidth]{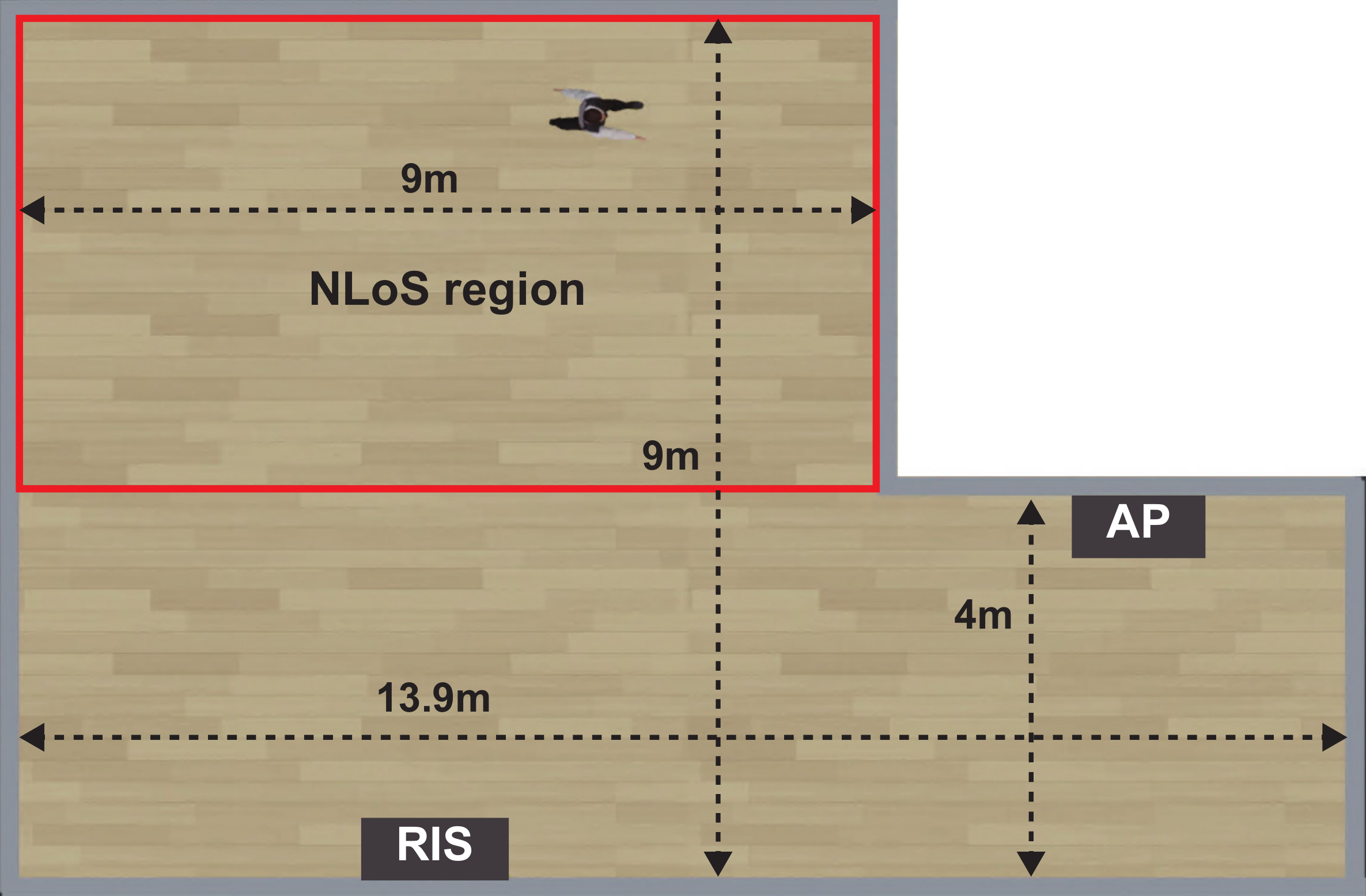}
		\caption{This figure depicts the adopted indoor scenario. We place the user in the NLoS area (red rectangle) to ensure that there is no direct link between the AP and the UE.}
		\label{fig:ray-tracing}
	\end{figure}

	\subsection{Simulation Framework}
	In this paper, we propose to leverage depth estimation to detect the user and design an RIS interaction vector for communication. It is essential to use realistic channels in the simulation since the sensing and communication performance highly relies on the environment geometry, scatterers' materials, etc. To that end, we follow the simulation framework in~\cite{taha2022} to generate the channels by accurate ray-tracing. Specifically, we first utilize Blender~\cite{Blender}, a high-fidelity 3D graphics design engine, to build a floor plan with a sufficient number of facets. Then, we export the designed floor plan to an accurate 3D ray-tracing simulator, Wireless Insite~\cite{Remcom}.

	For the RIS, we adopt a $40\times40$ uniform planar array structure at the mmWave 60 GHz operating band. The radar cross section gain of the RIS elements is assumed to be an isotropic gain with half-wavelength RIS element spacing. For the RIS-aided depth estimation, we generate the sensing channel paths and construct the received baseband digital signals \eqref{eq:z}. We consider an RIS sensing codebook with oversampling factors of four in both vertical and horizontal dimensions, i.e. resolution of $160\times160$ pixels. The configurations of the FMCW radar follow the settings in~\cite{taha2022}. For the RIS-aided communication, we generate the channel between the AP and the RIS, $\bh_T$, and the channel between the RIS and the UE, $\bh_R$. Then, the composite channel $\bh_T \odot \bh_R$ can be calculated.

	\subsection{Results for An Indoor Scenario}
	In~\figref{fig:ray-tracing}, we present the top view of the adopted scenario and mark the locations of the AP and the RIS. We consider an L-shape indoor space, where a 1.8m tall person is standing in the NLoS area. The materials of the objects/surfaces, including the concrete walls, floorboard, and ceiling board, are set to the ITU default parameter values at 60 GHz. We generate the samples by placing the person model at 32 distinct locations in the NLOS region, with equal spacing.   For each user location, we estimate the depth map and apply the user detection solution as described in Section \ref{sec:user_detect}. Further, to consider practical scenarios, for each user location, we place the UE's receive antenna at either the jacket pocket or the front/back pants pocket on the person model, resulting in a total of 96 generated samples. We adopt a classical beamsteering codebook to evaluate the performance of the proposed RIS beam selection approach. For comparison, we generate a beamsteering codebook with oversampling factors of four in both elevation and azimuth dimensions. 

\begin{figure}[!t]
		\centering
		\includegraphics[width=1\columnwidth]{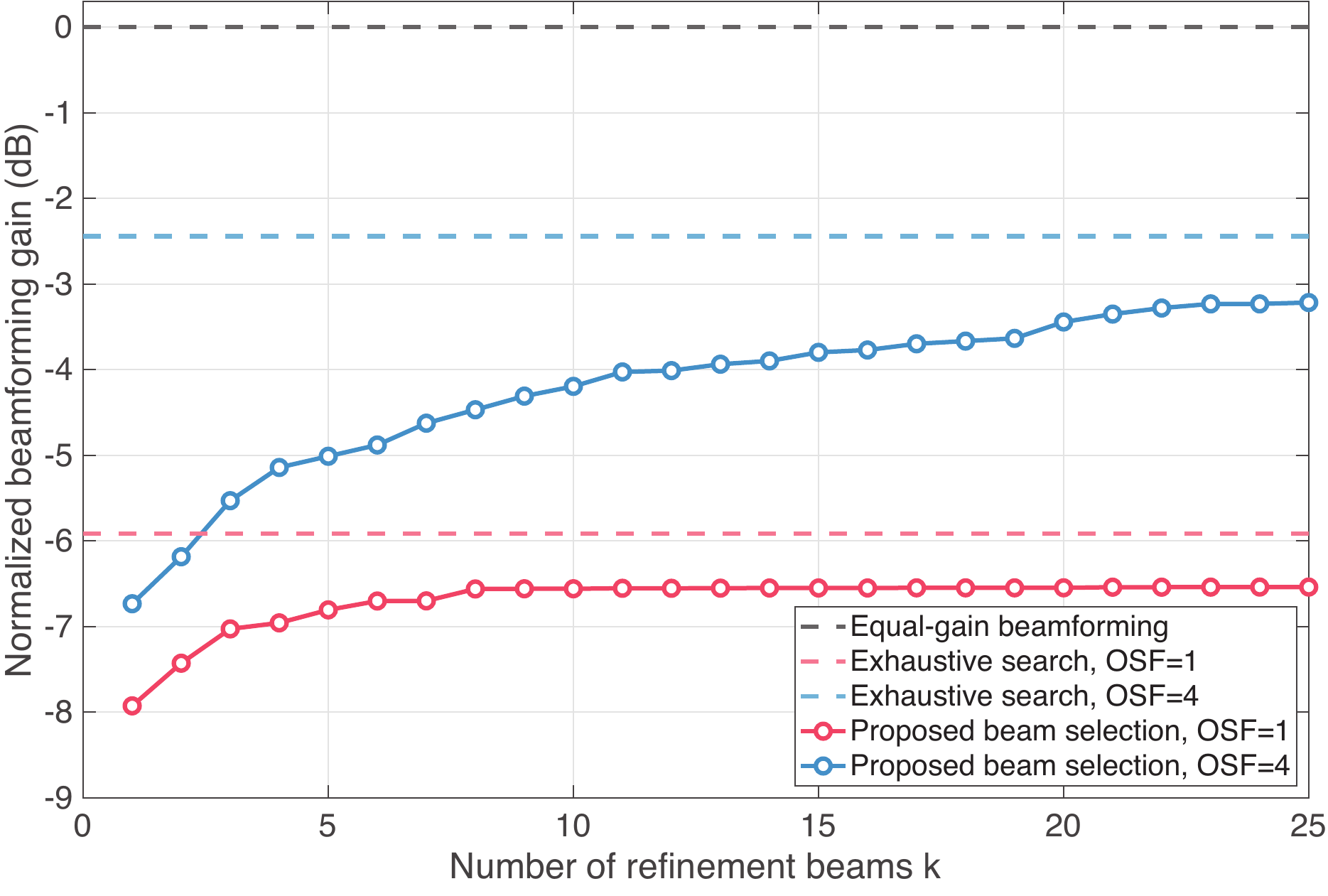}
		\caption{This figure presents the beamforming gain provided by the selected top-k beams, compared to the equal-gain beamforming (upperbound) and the exhaustive search. The oversampled codebook is generated with the oversampling factors (OSFs) of four in azimuth and elevation dimensions.}
		\label{fig:beamforming_gain}
	\end{figure}

	\figref{fig:beamforming_gain} shows the top-k normalized beamforming gain of the RIS beam selection. The selection of top-k beams is determined based on the similarities defined in~\eqref{eq:sim}. Specifically, we sort the codebook beams based on the similarity values in a descending order to find the top-k beams. The normalized beamforming gain is calculated as the ratio of beamforming gain to  the equal-gain beamforming gain. Note that, the equal-gain beamforming is assumed to have perfect channel knowledge. As the value of $k$ increases, better beamforming gain can be achieved with the proposed RIS beam selection approach. In particular, \textbf{with the oversampled codebook, the proposed solutions have comparable performance to the optimal beam of the codebook, while requiring less than $0.1\%$ of the beam training overhead of the exhaustive search, which demands $25600$ iterations.} Compared to the codebook without oversampling, the oversampled codebook yields a $3$ dB gain for the top-25 selected beams, which implies that better received SNR can be attained. This is because the RIS with a large number of elements enables narrow reflected beams, and the non-oversampled codebook can not provide high beamforming gain in all directions. In short, \textbf{the simulation results demonstrate the potential of using depth estimation to assist the RIS interaction vector design in the integrated imaging and communication system.}

	\section{Conclusion} \label{sec:con}
	In this paper, we investigate an RIS-aided integrated imaging and communication system that leverages scene depth estimation to achieve low RIS beam training overhead for communication. Specifically, we propose a user detection algorithm to position the user in an estimated depth map, which can be used in the RIS interaction vector design. Then, we design an RIS beam selection scheme based on a pre-defined codebook for communication. Simulation results reveal that the proposed solutions can overcome the large search space of the RIS interaction codebook with high beamforming gain and low overhead. This demonstrates the potential of imaging-aided communication in the proposed ISAC system.

	\balance


\begin{thebibliography}{10}
		\providecommand{\url}[1]{#1}
		\csname url@samestyle\endcsname
		\providecommand{\newblock}{\relax}
		\providecommand{\bibinfo}[2]{#2}
		\providecommand{\BIBentrySTDinterwordspacing}{\spaceskip=0pt\relax}
		\providecommand{\BIBentryALTinterwordstretchfactor}{4}
		\providecommand{\BIBentryALTinterwordspacing}{\spaceskip=\fontdimen2\font plus
		\BIBentryALTinterwordstretchfactor\fontdimen3\font minus
		  \fontdimen4\font\relax}
		\providecommand{\BIBforeignlanguage}[2]{{%
		\expandafter\ifx\csname l@#1\endcsname\relax
		\typeout{** WARNING: IEEEtran.bst: No hyphenation pattern has been}%
		\typeout{** loaded for the language `#1'. Using the pattern for}%
		\typeout{** the default language instead.}%
		\else
		\language=\csname l@#1\endcsname
		\fi
		#2}}
		\providecommand{\BIBdecl}{\relax}
		\BIBdecl
		
		\bibitem{demirhan2023}
		U.~Demirhan and A.~Alkhateeb, ``{Integrated Sensing and Communication for 6G:
		  Ten Key Machine Learning Roles},'' \emph{IEEE Communications Magazine},
		  vol.~61, no.~5, pp. 113--119, 2023.
		
		\bibitem{jiang2022a}
		Z.-M. Jiang, M.~Rihan, P.~Zhang, L.~Huang \emph{et~al.}, ``{Intelligent
		  Reflecting Surface Aided Dual-Function Radar and Communication System},''
		  \emph{IEEE Systems Journal}, vol.~16, no.~1, pp. 475--486, 2022.
		
		\bibitem{he2022}
		Y.~He, Y.~Cai, H.~Mao, and G.~Yu, ``{RIS-Assisted Communication Radar
		  Coexistence: Joint Beamforming Design and Analysis},'' \emph{IEEE J. Sel.
		  Areas Commun.}, vol.~40, no.~7, pp. 2131--2145, 2022.
		
		\bibitem{taha2022}
		A.~Taha, H.~Luo, and A.~Alkhateeb, ``{Reconfigurable Intelligent Surface Aided
		  Wireless Sensing for Scene Depth Estimation},'' in \emph{IEEE International
		  Conference on Communications}, 2023, pp. 491--497.
		
		\bibitem{ramasubramanian2017}
		K.~Ramasubramanian and T.~Instruments, ``{Using a Complex-Baseband Architecture
		  in FMCW Radar Systems},'' vol.~19, 2017.
		
		\bibitem{li2021}
		X.~Li, X.~Wang, Q.~Yang, and S.~Fu, ``Signal processing for {TDM MIMO FMCW}
		  millimeter-wave radar sensors,'' \emph{IEEE Access}, vol.~9, pp.
		  167\,959--167\,971, 2021.
		
		\bibitem{ester1996}
		M.~Ester, H.-P. Kriegel, J.~Sander, X.~Xu \emph{et~al.}, ``{A Density-Based
		  Algorithm for Discovering Clusters in Large Spatial Databases with Noise},''
		  in \emph{Proc. of the 2nd Int. Conf. on Knowledge Discovery and Data Mining},
		  vol.~96, no.~34, 1996, pp. 226--231.
		
		\bibitem{jiang2022b}
		S.~Jiang, A.~Hindy, and A.~Alkhateeb, ``{Sensing Aided Reconfigurable
		  Intelligent Surfaces for 3GPP 5G Transparent Operation},'' \emph{IEEE
		  Transactions on Communications}, vol.~71, no.~11, pp. 6348--6362, 2023.
		
		\bibitem{Blender}
		\BIBentryALTinterwordspacing
		B.~O. Community, \emph{{Blender - A 3D Modelling and Rendering Package}},
		  Blender Foundation. [Online]. Available: \url{http://www.blender.org}
		\BIBentrySTDinterwordspacing
		
		\bibitem{Remcom}
		Remcom, ``{Wireless InSite},'' \url{http://www.remcom.com/wireless-insite}.
		
	\end{thebibliography}
\end{document}